\documentstyle[aps,preprint]{revtex}
\newcommand{\be}{\begin{eqnarray}}
\newcommand{\ee}{\end{eqnarray}}

\begin{document}
\pagestyle{plain}
\newcount\eLiNe\eLiNe=\inputlineno\advance\eLiNe by -1

\title{ FLUCTUATIONS OF WIG-\\ 
        the index of Warsaw Stock Exchange. \\
         Preliminary studies
\thanks{Presented at 2000 Marian Smoluchowski Symposium on Statistical Physics, Zakopane, Poland, September 10--17,2000} }
\author{Danuta Makowiec and Piotr Gnaci\'nski\\
{Institute of Theoretical Physics and Astrophysics, \\
 Gda\'nsk University, Poland, {\it fizdm@univ.gda.pl } }
}
\maketitle
\begin{abstract}
A time series that represents daily values of the WIG index (the main index of Warsaw Stock Exchange) over last 5 years  is examined. 
Non-Gaussian features of  distributions of fluctuations, namely returns, over a time scale are considered.  
Some general properties like exponents of the long range correlation  estimated by  averaged volatility and detrended fluctuations analysis (DFA) as well as  exponents describing a decay of tails of the cumulative  distributions are found. Closing, the Zipf analysis for  the WIG index  time series translated into three letter text is presented.
\end{abstract}
{\bf PACS{ 05.40.Fb, 05.45.Tp, 89.90.+n}}
\section{Introduction}
Financial markets are complex dynamical systems  with many interacting elements: individual investors, mutual funds, brokerage firms, banks from one side and  bonds, stocks, futures, options from the other side. Interactions between elements lead to transactions mediated by the stock exchange. The details of each transaction are recorded for analysis.
The nature of interactions between different elements comprising a financial market is unknown as well as the way in which external factors affect the market. Therefore, as a starting point  one may turn to empirical studies to inquire into regularities called  "empirical laws" that may drive financial markets.

In the following we present results of investigations of the WIG index -
the index of Warsaw Stock Exchange. (WIG is an abbreviation of  the Polish name.) The WIG index is calculated as a total return of the weighted sum of market capitalization (a portfolio of the index) of all stocks from the main market once per trading day after each session. Every three months the portfolio is revised. New stocks that entered the market in previous three months  are added and the portfolio weights  are corrected to preserve two rules: any market capitalization of a stock takes more than 10\%   and participation of one industry stocks do not exceed 30\% of the whole market capitalization.  The Warsaw Stock Exchange is a young market--- the first session took place on the 16th April of 1991 and five stocks were traded. Now, i.e.  on 1 August 2000 the portfolio of the WIG index enumerates 120 stocks of the main market and 61 stocks are traded on the parallel market. 

In this presentation we study a time series of returns of the WIG index for the period of 5 last years: from September 1995 to August 2000. Thus we observe the emerging market in its second phase of development. Therefore we can assume that the market is stationary  (see \cite{Budapest} for emerging market studies).

The presentation is organized as follows: 

In Section 2, we introduce a time series considered and give a definition for returns- one of the basic notions in a study of financial markets.

 In Section 3, we perform  tests to estimate strength and character of long-range correlation of Warsaw market. We find  an instance of the  strongly antypersistant random walk occurring in the time horizon longer than three months. 

In Section 4, we discuss properties of probability  density functions of returns. We show that the central part of distribution of WIG index returns is well fitted by a L\'evy distribution. Some time scaling is therefore provided. The asymptotic behavior of the distribution of returns shows faster decay than predicted by a L\'evy distribution. Hence, our result confirms Mantegna and Stanley proposition \cite{M&SNature, M&SBook} of a truncated L\'evy distribution as a model for the distribution of returns. The exponential truncation ensures the existence of a finite second moment what concludes, by limit theorems, that the asymptotic distribution of returns is a Gaussian distribution.

The next section, Section 5, we translate the time series of WIG index values into a text \cite{DNA,Vandewalle}. The Zipf \cite{Zipf,Vandewalle} analysis drives to the observation that non trivial correlation exist between successive daily fluctuations, so that some predictions are possible.

\section{Data in study}
We analyze the values of the WIG index for the period of almost 5 years: from 1st September 1995 to 31st July 2000. The data have been recorded at each trading day. So that our basic data set consists of 1224 values.
In Fig.1 we present a time series corresponding to daily values of the WIG index. Together, we plot the WIG index detrended by the inflation rate. Because of high variety in time it is difficult to give an estimation of satisfactory quality  for the total trend. Rough calculations for the series of the detrended WIG index suggest the trend is neither upwards nor downwards. The total trend over the last five years is horizontal.

To test the nature of the stochastic process underlying the changes in the WIG index value we investigate the time series of returns on varying time scales $\Delta t$. Returns are the basic quantities that are widely  studied in economic analysis, see \cite{M&SBook,Bouchaud,Gopikrishnan99} for a literature collection. For a time series $S(t)$ of prices or market index values, a time series of {\bf returns }  $G({\Delta t}, t)$ over  a time scale $\Delta t$ 
is obtained as the successive differences of the natural logarithm  of price, 
\cite{Gopikrishnan99}
\be
G (\Delta t, t) = \ln S(t+\Delta t) - \ln S(t)
\ee
A normalized to the variance 1 series of returns $g(\Delta t, t)$ is
\be
g({\Delta t}, t) = {G({\Delta t},t ) -<G({\Delta t }, t)>_T \over v(\Delta t) }
\ee
where $v^2(\Delta t)$ is the time averaged variance,
\be
v^2(\Delta t)= <G^2({\Delta t},t )>_T - < G({\Delta t}, t )>_T^2 
\nonumber \ee
and $<..>_T$ denotes an average over the entire length of a time series.
A daily series of $g(1,t)$ together with examples of time series when the data set is sampled at $ 5, 10$ and $20$ trading days are shown in Fig.2.

\section{Long-time correlation}
A commonly used and efficient test in detecting the presence of long-range correlation is based on investigation the standard deviation of returns as a function of different time horizon $\Delta t$. This function is called the time averaged {\bf volatility} and is denoted $v(\Delta t)$, \cite{Gopikrishnan99}. (In economics, the volatility of a certain stock is a measure 
how much a price is likely to fluctuate at a given time. It can also be related to the amount of information arriving at any time. Such an instantaneous volatility can be calculated as the local average of the amplitude of the returns, \cite{M&SBook}.)

The volatility dependence  on the time scale $\Delta t$
allows to determine an exponent $\delta $ of a power-law behavior
\be
 v(\Delta t) \sim \Delta t^\delta  .
\ee
For a random walk $\delta =0.5$. In mature markets the exponent $ \delta$ takes value larger than but close to $0.5$ what indicates absence of correlation for the daily and monthly returns \cite{Gopikrishnan99} or a walk with a weak persistent character \cite{M&SBook,Bouchaud}.

Fig.3a is the log-log plot of the  volatility of the WIG index returns against time  to extract the exponent $\delta $. It shows that initially for two to five days, the exponent takes the value  $\delta = 0.553 $ what implies a slightly persistent walk. Then, for a time period of few further days, the exponent switches to  $\delta = 0.464$ what can be recognized as slightly antypersistent (zigzag) walk. If longer intervals are considered such as one or two month time, the volatility  grows with  $\delta = 0.557 $. (In brackets the Pearson $r^2$ correlation coefficient is shown to give the precision of  the linear dependence.)
Finally, on the time scale of three or more months, one  observes $\delta=0.269$(!).

For verification the last value we have performed a similar analysis for a series of returns obtained from the data of  other indices of the Warsaw Stock Exchange and some particular stocks. These results are shown in Figs 3b,c,d. The WIG-20 (twenty largest firms according to market capitalization) index exhibits the same features as the WIG index. The WIRR index (the index of the parallel market) and the prices of some particular stocks do not change the character of correlation on the three-month scale. Instead, we observe the amplification of the non-random walk features.  One can notice the presence of strong long-range correlation in the most of the time series. 
We think that the rapid change in the trend of fluctuations of the WIG and WIG20 indices is effected by corrections made in calculating procedure for these indices. Every three months the WIG and WIG20 indices formulas  are changed.

In Fig.4 one can find  a plot of the exponent $\alpha$ arising from the Detrended Fluctuation Analysis (DFA) method \cite{DFA}. Again, one can notice switching from the persistent random walk taking place in the first week time to antypersistant walk emerging in the second week. The growing errors of regression fits, measured by the Pearson $r^2$ correlation coefficient, do not allow us to present $\alpha$s  for returns over  the longer time horizon.

\section{Probability density function}
One of the key aspect in determining stochastic evolution of the value of a given financial asset is the shape of the probability density function (pdf) of its returns. Fig.5a shows on the semi-log plot the pdf of one-day returns $G(\Delta t=1)$. Searching for properties of experimental pdf of returns we fit the data to the Gaussian and  Lorentzian distributions.

The rough analysis of Gaussianness of an empirical pdf can be done by considering the {\bf kurtosis } --- the fourth normalized cumulant. The kurtosis equals to 0 for a Gaussian distribution. When the kurtosis is positive --- {\bf a leptokurtic distribution}, the corresponding distribution density has a market peak around the mean and rather 'thick' tails. Conversely, when the kurtosis is negative, the distribution density has a flat top and very thin tails.  Therefore, the kurtosis is considered as a measure of a distance between an empirical data distribution and a Gaussian curve. The leptokurtic character of a stock price returns as well as stock index returns has been reported by several authors since Mandelbrot observation in 60-ties, \cite{Mandelbrot}.

Fig.5b shows values of the kurtosis for returns over the time scale.
The leptokurtic character of distributions of the WIG index returns is evident when the distribution of returns over a time scale shorter than two weeks (10 trading days) is considered.  Over a time scale of two, three months  one should observe the change of the pdf shape to the opposite flat character.  

In Fig.6a the probability density functions for returns, namely $G(\Delta t )$, over different time scales $\Delta t$ are shown. The typical spreading is observed,  as if a random walk is examined. Having disagreement between the empirical pdf and a Gaussian distribution though supposing  stability of distributions, we search for a symmetrical L\'evy distribution,
\be
P_\mu (x)={1\over \pi} \int_0^\infty e^{-\gamma |q|^\mu} \cos(qx) dx
\ee
To find the most appropriate stable L\'evy pdf we calculate the probability of return to the origin, i.e., Prob$\{ G(\Delta t) = 0 \}$, as it is usually proposed \cite{M&SBook,Gopikrishnan99}.
Plotting Prob$\{ G(\Delta t) = 0 \}$ against different time scales $\Delta t$ on the log-log plot, we can observe two time regimes with slightly different power-law behavior, see Fig.6b. The returns of few days time scale lead to the exponent value $0.676$ while the returns of few weeks time scale provide $0.578$ for the exponent. The index $\mu$ of a L\'evy distribution is the inverse of the exponent and  thereby we get $\mu_{days}= 1.470$ and $\mu_{weeks}=1.730$. Both values are well inside the L\'evy stable range $0\le \mu <2$. It is known that the value of $\mu$  for stock markets falls between 1.5 and 1.7 and  is lower for more volatile markets \cite{Cont,M&SBook} and on the time scales longer than few days one should expect a  convergence to Gaussian behavior \cite{Gopikrishnan99}. Hence, according to our results the Warsaw stock market is highly volatile and this volatility slows down the convergence to Gaussian behavior.

The stable distributions are self-similar --- the distributions of returns for various choices of $\Delta t$ have the similar functional form. Having value of  $\mu$, we rescale our empirical pdfs  to observe data collapse, see Fig.6c.
Applying suitable  scaling to different pdfs we find two separated areas with collapsing data. One area corresponds to returns with  few days time scale and the other to returns of few weeks time horizon.

Searching for the breakdown of a L\'evy description in  empirical pdfs,  
rare events are usually  considered. The probability of rare events are estimated by studying properties of the  cumulative distribution of the normalized returns $g(t)$ \cite{M&SBook,Gopikrishnan99}.
The cumulative probability distribution $F(g)$ of observing a change $g$ or larger in a series of returns was found to be power-law for large values of $g$, both for positive  and negative values of $g$ with exponent $\mu $,
\be
F(g)=Prob\{ g > x \} \sim g^{-\mu}
\ee
what  implies the following asymptotic behavior of pdfs,
\be
P(x) \sim {1\over x^{1+\mu}},  \qquad |x| \quad {\rm large \ enough}
\nonumber \ee

The value of $\mu$ changes when one moves the interval of $g$. Our estimates are, Figs 7ab:
\be
{\rm for} \quad g\in(0.3,0.9): \qquad  \mu=\cases{0.76 &positive tail\cr
                              0.69 & negative tail }\nonumber \ee
\be
{\rm for} \quad g\in(0.9, 1.6): \qquad  \mu=\cases{2.03 &positive tail\cr
                              1.83 & negative tail }
\nonumber \ee
\be
{\rm for} \quad g > 1.6  : \qquad  \mu=\cases{3.88 &positive tail\cr
                              3.06 & negative tail }
\nonumber \ee

The presented  results are consistent with the apparent power-law behavior in tails found for daily returns for stock market indices such as NIKKEI: $\mu \approx 3.05$, Hang-Seng : $\mu \approx 3.03$ or S\&P500 : $\mu \approx 3.34$, \cite{Plerou,Gopikrishnan99}.
Moreover, since $\mu > 2$ when $g> 1.6$, hence it is outside the L\'evy stable  regime, our investigations lead to the conclusion that the second moment of fluctuations of the WIG index is finite and one should expect the convergence to a Gaussian distribution because of the central limit theorem. 
One should notice that the value for $\mu$ found by study the asymptotic nature of the distribution is different from the value obtained when the center part of the distribution is considered. This effect is usually related to the slow decay of the volatility correlation \cite{Gopikrishnan99,Liu}.

\section{Zipf plot of WIG index}
The original Zipf analysis \cite{Zipf} consists in counting the number of words of a certain type appearing in a text, calculating the frequency of occurrence $f$ of each word in a given text, and sorting out the words according to their frequency, i.e., a rank $R$ is assigned to each word, namely, $R =1 $ for the most frequent one. A power law
\be
f\sim R^{-\zeta }
\ee
with an exponent $\zeta $ is searched for on a log-log plot.

The similar analysis can be done to any series of data when one translates the data into a text \cite{DNA}. In the following we translated the WIG index time series into a text based on two alphabets. The first text is obtained by transforming  data into binary sequences by replacing up and down daily fluctuations by a character $u$ and $d$,  respectively.
Another transformation: $u$ for large up $d$ for large down fluctuations and $s$ for small fluctuations of both kinds below a given threshold, leads to a three-letter text.

In Fig.8 we show the Zipf plot for words consisting of $n= 4,5$ letters of both texts obtained from the time series of WIG index values.
The probability of a single letter to be found in the text examined is as follows:
$P(u)=0.533$ and $ P(d)=0.467$ when the two-letter alphabet is considered and 
if we deal with three letter alphabet then depending on the value of the threshold $s$ 
we obtain: 
\be
\begin{array}{cccccc}
 s= & 2\% & 1\% & 0.6\% & 0.2\% & 0.1\% \cr
d&0.103 & 0.234 & 0.316 & 0.406 & 0.441 \cr
s & 0.774 &0.510 & 0.318 & 0.119 & 0.059 \cr
u & 0.123 & 0.255 & 0.366 & 0.475 & 0.500 \cr
\end{array}\nonumber
\ee
The data plotted are the effective frequencies $P\slash P'$, i.e.,  the apparent probability $P$ of a word divided by its expected probability of occurrence $P'$
if there is no correlation \cite{Vandewalle}. The Zipf plot using $P\slash P'$ is expected to be horizontal for random sequences.

The value of $\zeta$ changes depending on the way in which a time series is transferred into a text:

Two-letter alphabet:
\be
\begin{array}{cccccc}
{\rm 4-letter\quad words } & 0.15 \qquad(r^2=0.84) \cr
{\rm 5-letter\quad words } &0.18 \qquad (r^2=0.83) \cr
\end{array} \nonumber
\ee
Three-letter alphabet:

4-letter words:
\be
\begin{array}{cccccc}
 s & 2\% & 1\% & 0.6\% & 0.2\% & 0.1\% \cr
 \zeta & 0.86 &0.35 &0.24& 0.30 &0.64\cr
(r^2)&\quad(0.94) &\quad(0.98) &\quad(0.91)&\quad(0.98)&\quad(0.97)\cr
\end{array} \nonumber
\ee
5-letter words: 
\be
\begin{array}{cccccc}
\zeta &1.16 & 0.41 &0.26 &0.37 & 0.79\cr
(r^2)&\quad(0.95)&\quad(0.95)&\quad(0.91)&\quad(0.90)&\quad(0.98)\cr
\end{array} \nonumber
\ee
The lowest value is observed when probabilities of particular letters are approximately equal to each other and this happen when the bias $s=0.6\% $.
The value of $\zeta$ in this case is distinguishable different from zero, hence, non trivial correlation exist between successive daily fluctuations of the WIG index returns. Feeling tempted to make some predictions we search for no risk sequences of events, i.e., for the conditional probability equal to almost one that after a sequence of four events the fifth even occurs. We have found one of such an event: $Prob(d|ssud)\approx 1$ if $ s=0.6\% $ is considered. The set of such events strongly depends on the threshold value. Therefore we think that this analysis should be done for  price of a stock instead of market index.

\section{Conclusion}
It appears that some tools of statistical physics which were developed to study critical phenomena in complex systems can be succesfully applied in rather distant from physics disciplines like medicine, genetics, sociology, politics or economy. 
Many physicists have been attracted to interdisciplinary investigations collectively called exotic statistical physics \cite{Exotic}. 
Additional excitement comes along when one turns its interest into econophysics since one can learn not only how the economic system works  but also how to use this knowledge to make money \cite{Stauffer}. 
The authors of this presentation seem to be typical victims of such prospects.  
The main purpose of the presented paper is to apply standard analysis of experimental econophysics to Warsaw Stock Exchange index to discover how  to succeed on the Polish stock market.

Our conclusion that Warsaw Stock Exchange is volatile is not surprising. 
The Polish stock market, as it happens in all countries that change their economic systems, is small one, therefore, it is in an extremely fragile state.
The absence of a general positive trend in time series of the WIG index suggests that speculations are the only way to get an income or growth.
However, we have to emphasize that our conclusion is only preliminary.
The data set considered by us is rather limited; moreover, we do not explore and carry out all possible tests. 
We do not perform Hurst analysis of the data nor we study long-range dependences which emerge when the time dependent volatility is being investigated. 
The cross correlation between different stocks also has not been considered here.
Hence there is a need for further investigations to verify our initial findings.

{\bf Acknowledgment}\\
This work was partially supported by Gda\'nsk University grant BW 5400-5-0033-0

{\bf Figure captions}
\begin{itemize}
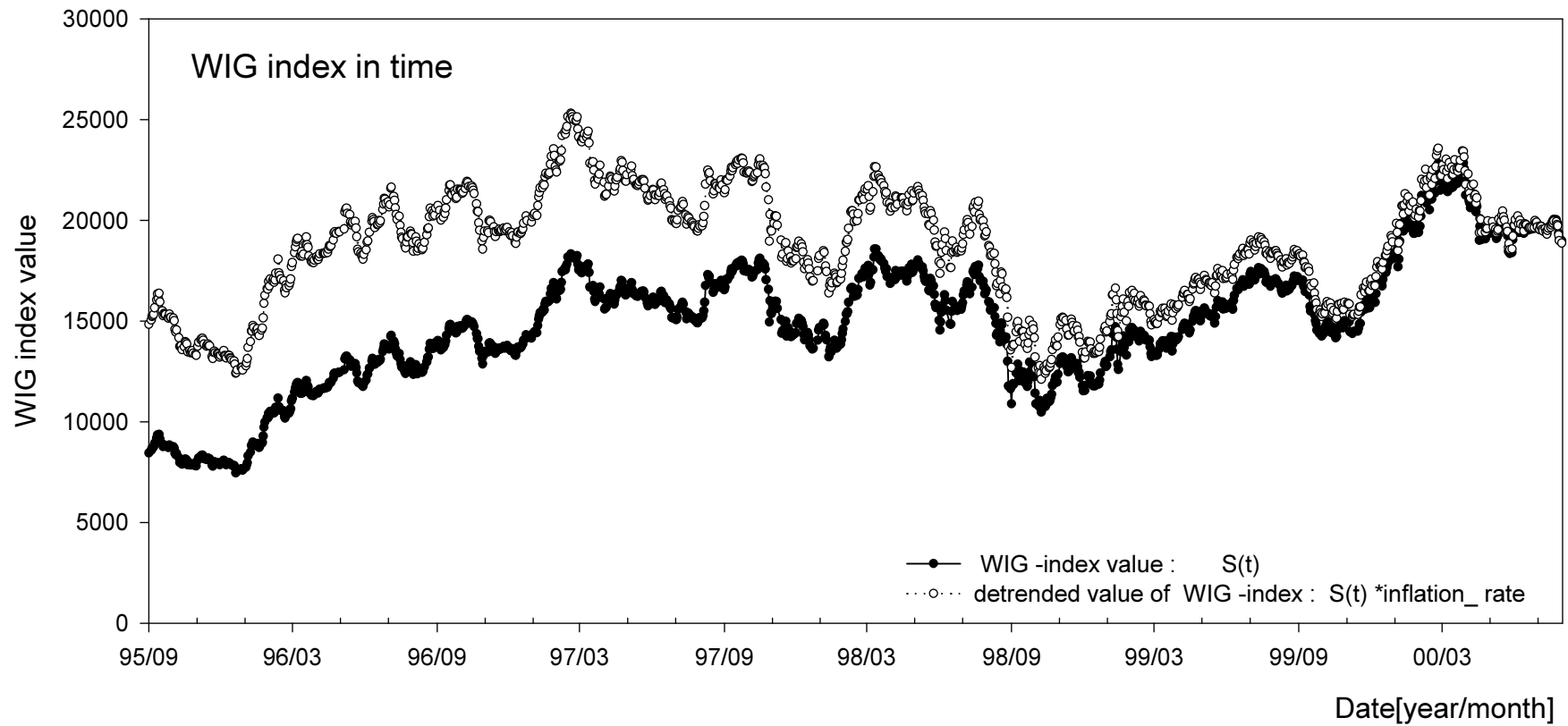
\item[Fig.1]Daily values of the WIG index in time. We display both the value of the WIG index (black dots) and the index detrended by inflation to the August 2000 of Polish zloty (empty dots).
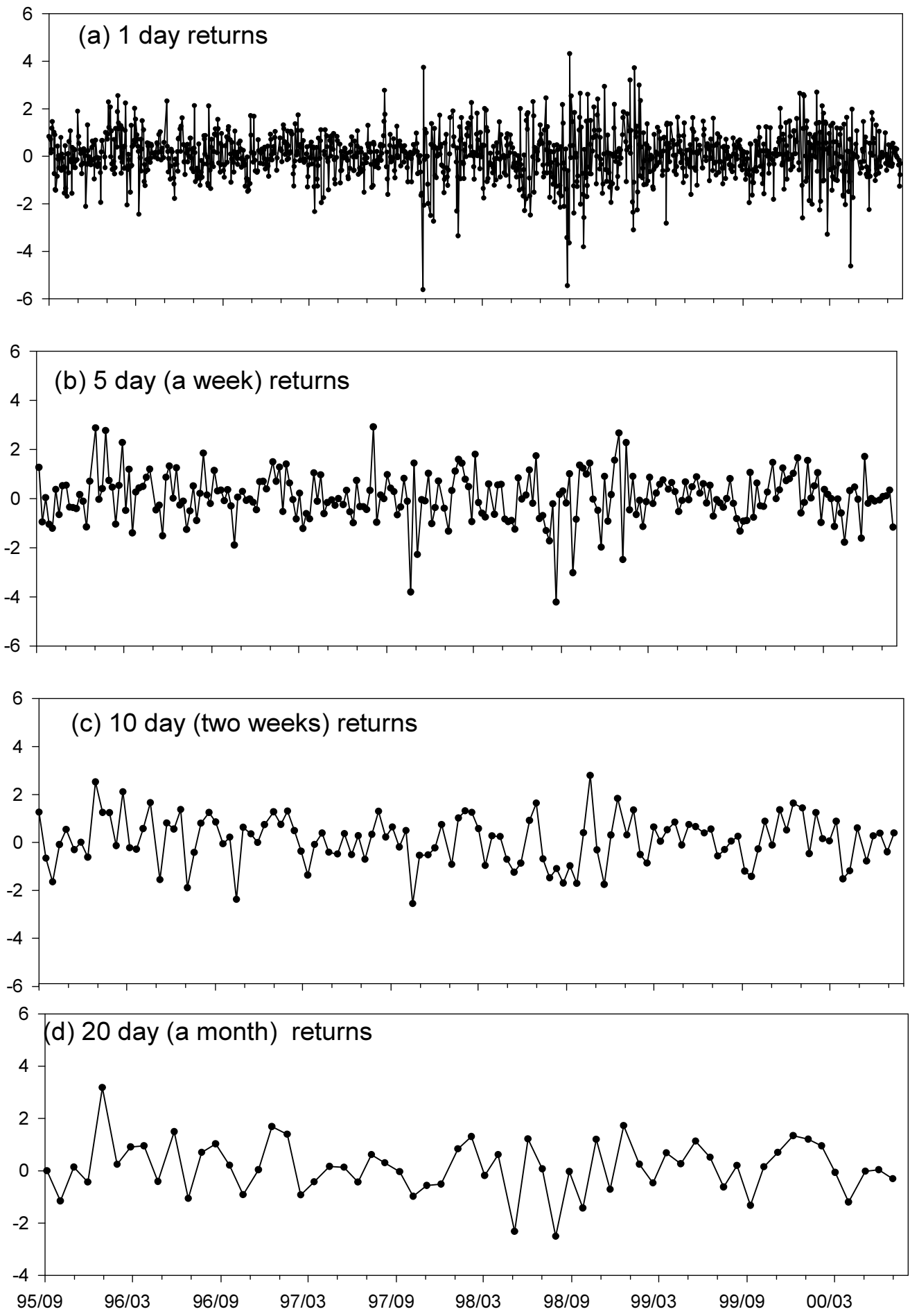
\item[Fig.2]Sequences of (a) one-day, (b) 5-day (a week), (c) 10-day, (d) 20-day (a month) returns extracted from the WIG index data.
\item[Fig.3]Log-log plot of the time averaged volatility as a function of the time scale $\Delta t$ of returns of (a) WIG index, (b)WIG20 index, (c) WIRR index (d) selected stocks. Notice, the rapid change of the correlation trend after three months in case the WIG and WIG-20 indices which is not observed when the WIRR index or particular stock prices are considered. 
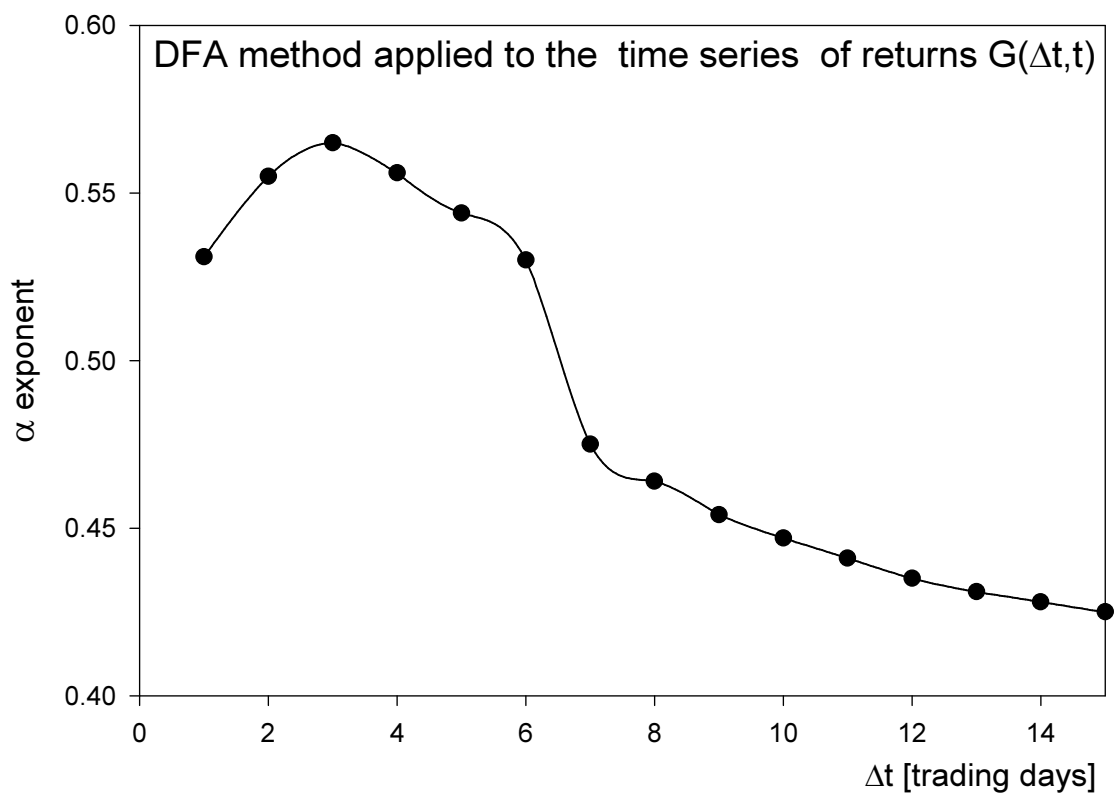
\item[Fig.4]Multifractality analysis by DFA method. The value of the index $\alpha$   is calculated for a time series that was created by merging one after one, all picked up series generated from our data set.  So that, we merge two time sequences of  G(2,t), three time sequences of G(3,t), etc., to deal with a time series of the reasonable length for DFA studies.
\item[Fig.5](a) Linear-log plot of the empirical probability density function for  $\Delta t=1 $ returns $G(1)$ of the WIG index. The best fit curve  to a Gaussian distribution is $26.7 \ \exp\{-0.5 ({x-0.0011\over 0.014 })^2\}.$
The best fit curve  to a Lorenztian distribution is ${30.1\over 1+ ({ x-0.0012)\over 0.012} )^2}$. 
(b) Kurtosis and skewness of the returns of WIG index. A positive value of kurtosis indicates a slow asymptotic decay. The 'leptokurtic' character of the distribution of the data implies that using a Gaussian one systematically underestimates the probability of large  fluctuations. 
The negative skewness says that a distribution is asymmetric around its mean and 
the negative tail of a distribution is larger than the positive one.
\item[Fig.6](a) Probability density function of returns $G(\Delta t )$
 measured at different time horizons: $\Delta t = 1, 2, 3$ and $10, 15, 21$ days.
(b) Probability of return to the origin measured as a function of the time interval $\Delta t$. A L\'evy distribution has a slope $\ge 0.5$, a Gaussian distribution has a  slope = 0.5. 
(c) Probability density functions of the returns plotted in rescaled units.
 Pdfs of short time scale returns are rescaled with $\mu_{days}$ and pdfs of returns of weeks time scale with $\mu_{weeks}$. 
\item[Fig.7](a)Log-log plot of the cumulative distribution of normalized WIG returns  for $\Delta t=1 $ day (negative tails). (b) Log-log plot of the cumulative distribution of normalized WIG returns  for $\Delta t=1$ day
(positive tails).
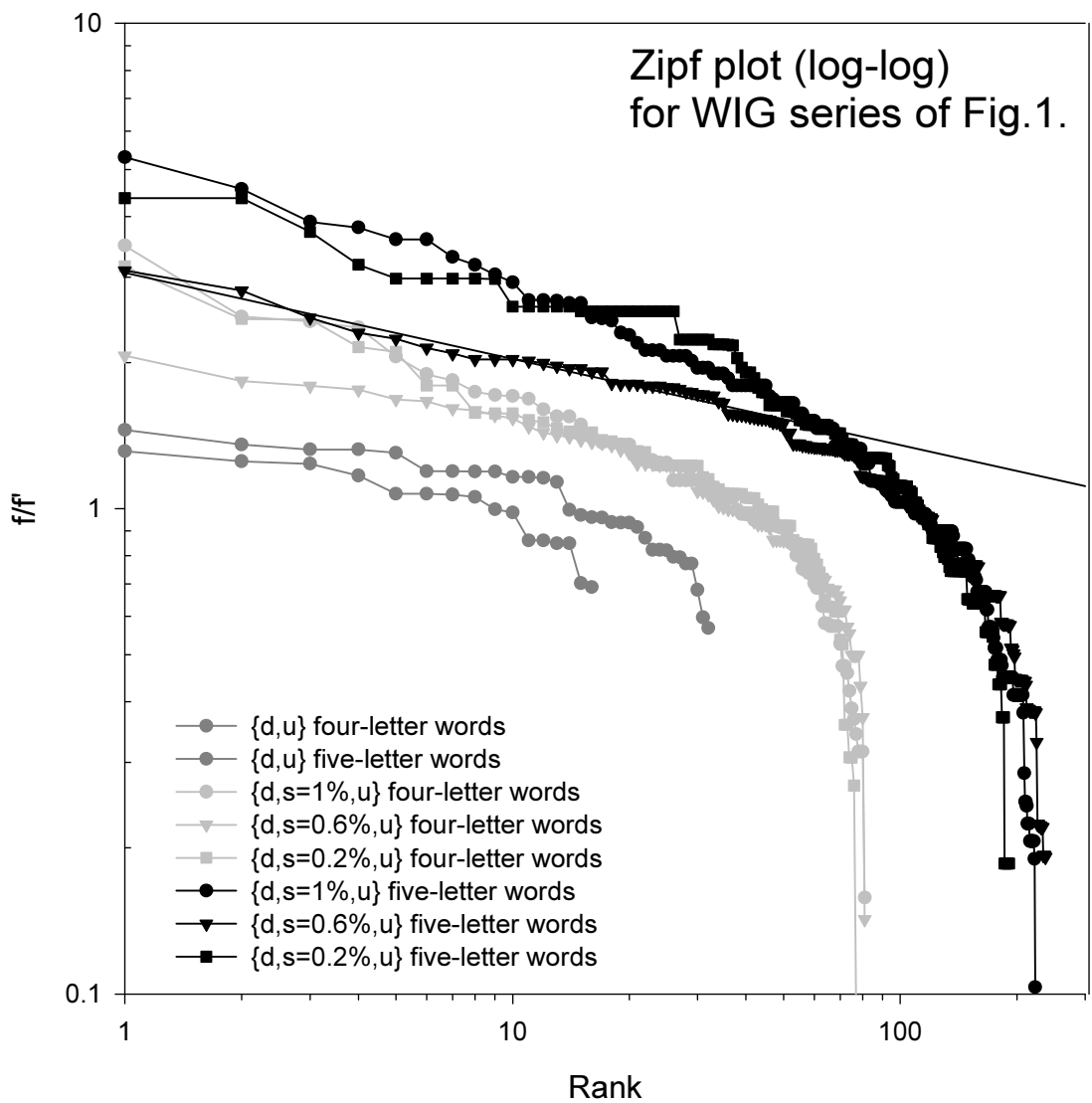
\item[Fig.8] Zipf plots for respectively the $u,d$ and $s,u,d$  sequences obtained  from the series of WIG returns of Fig.1. The length of analyzed pattern is $n=4, 5$.

\end{itemize}

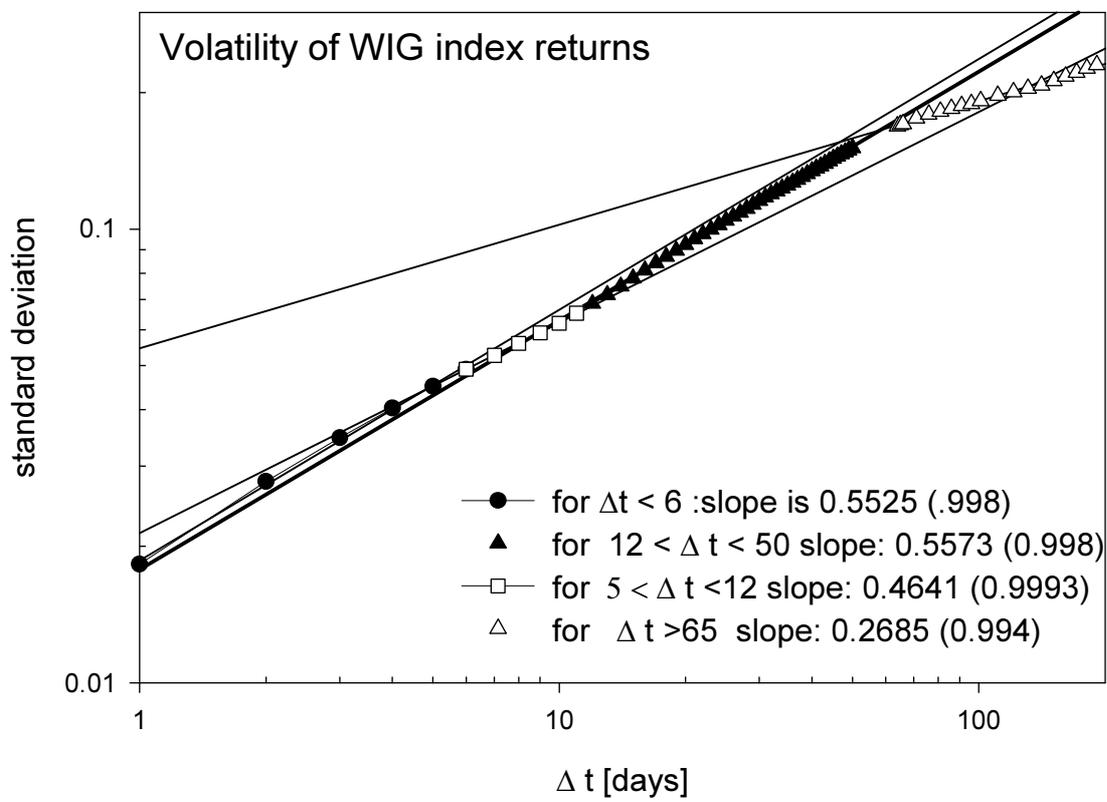
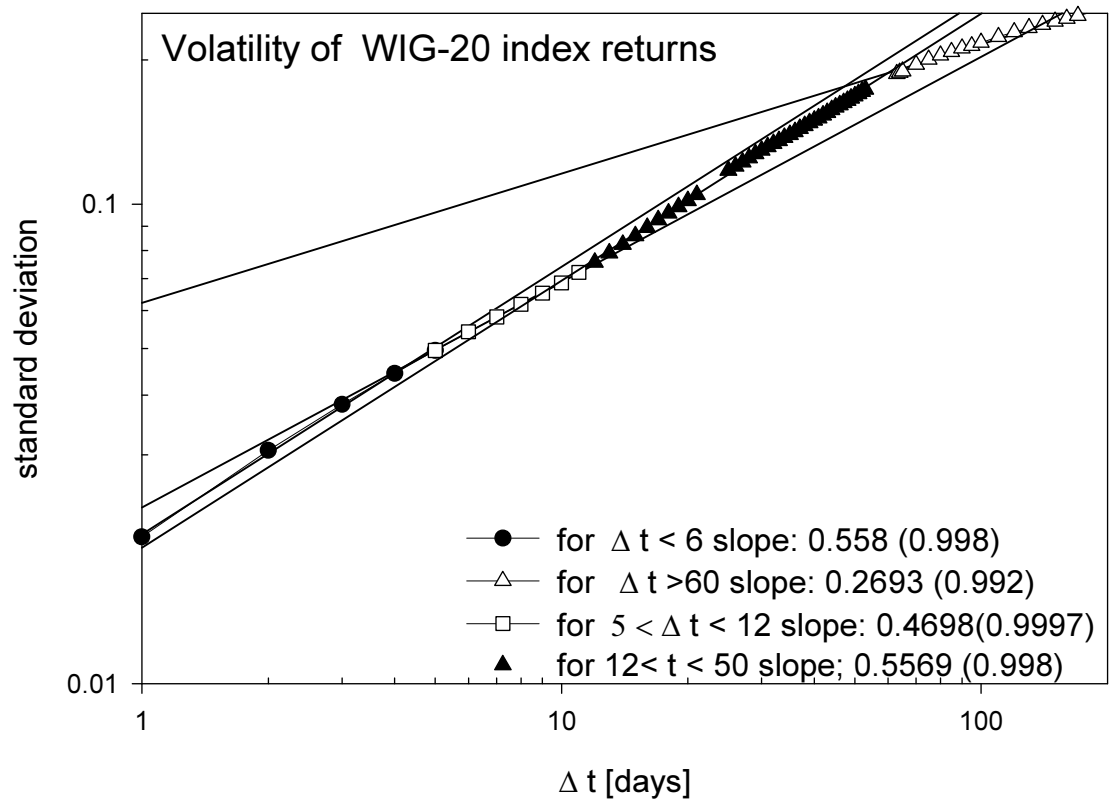
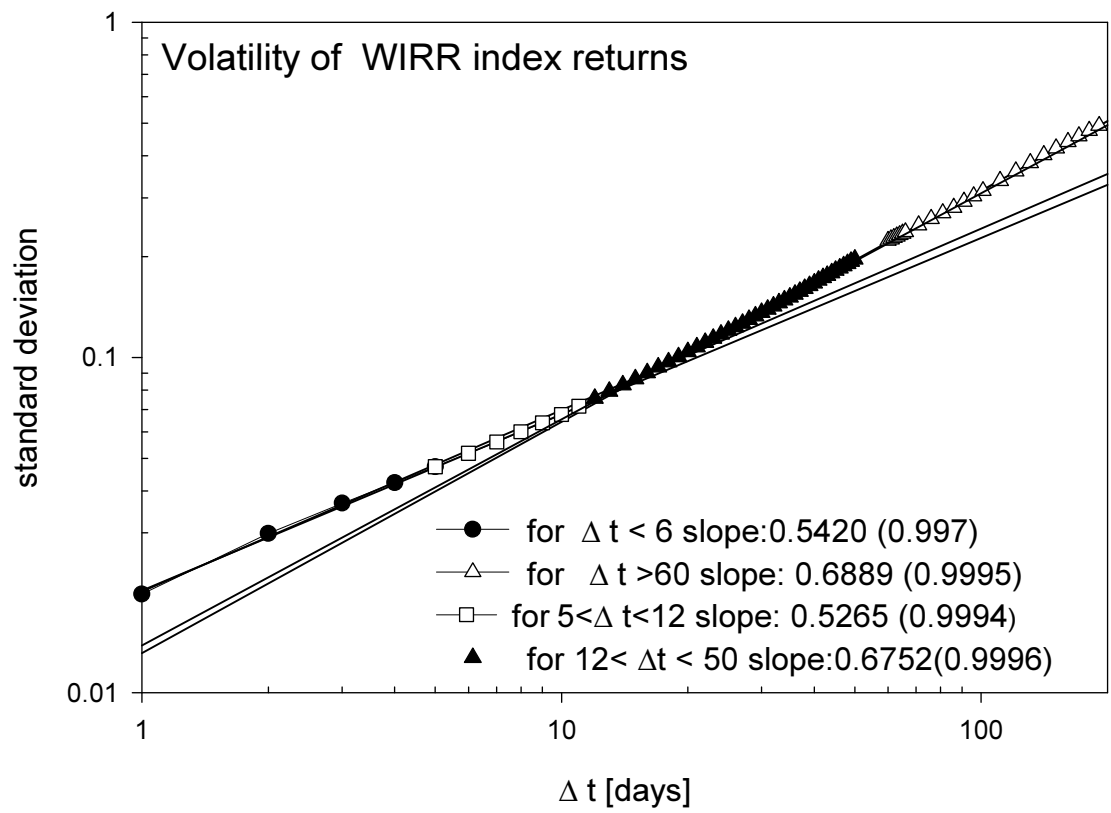
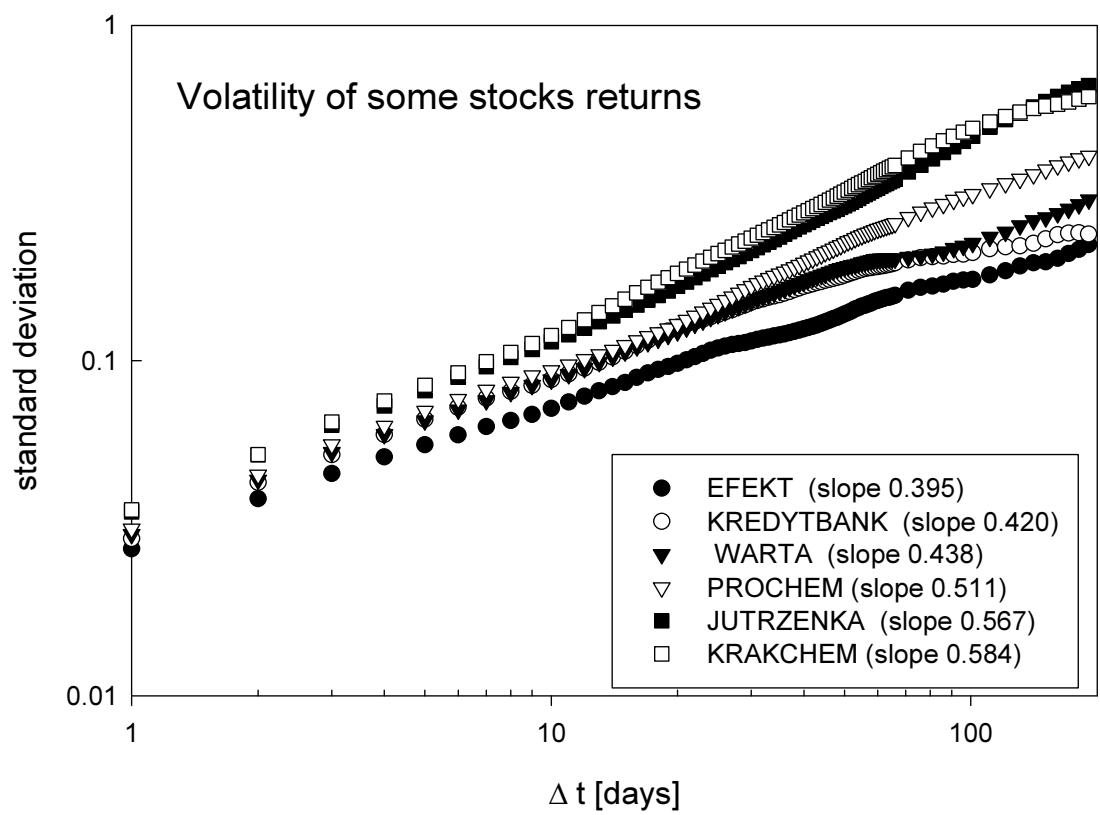
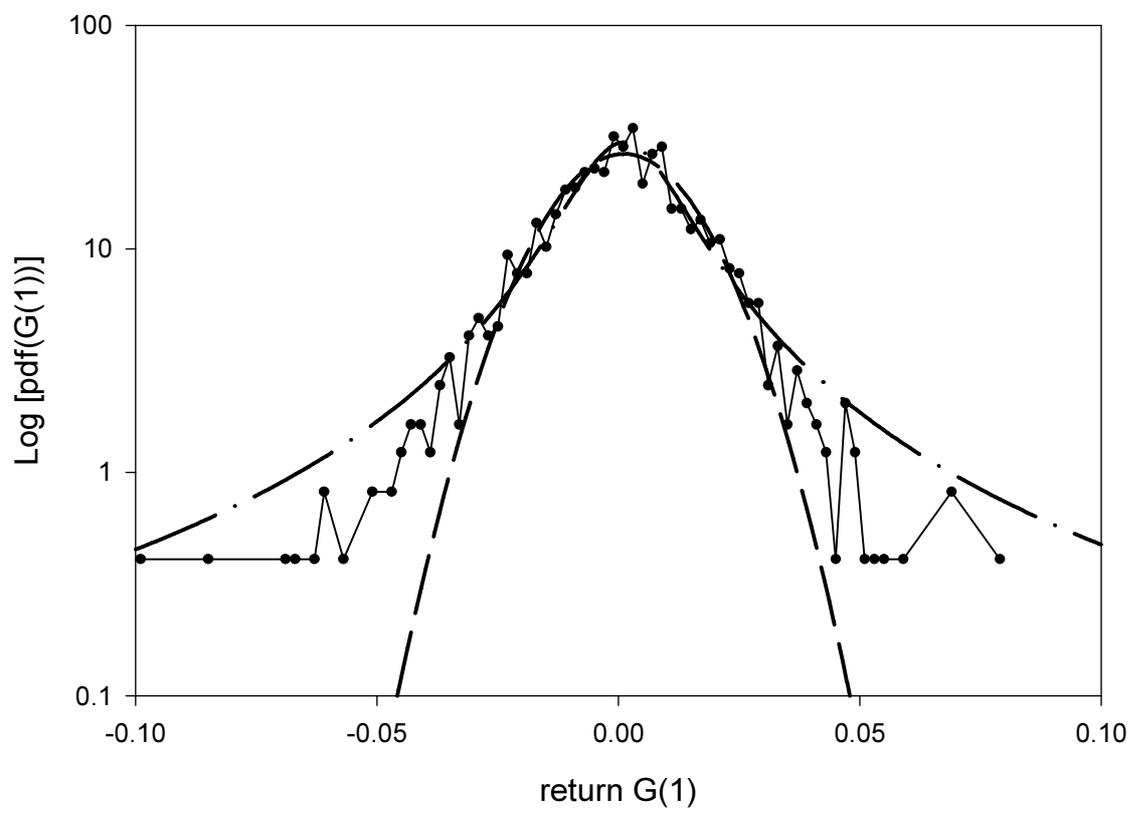
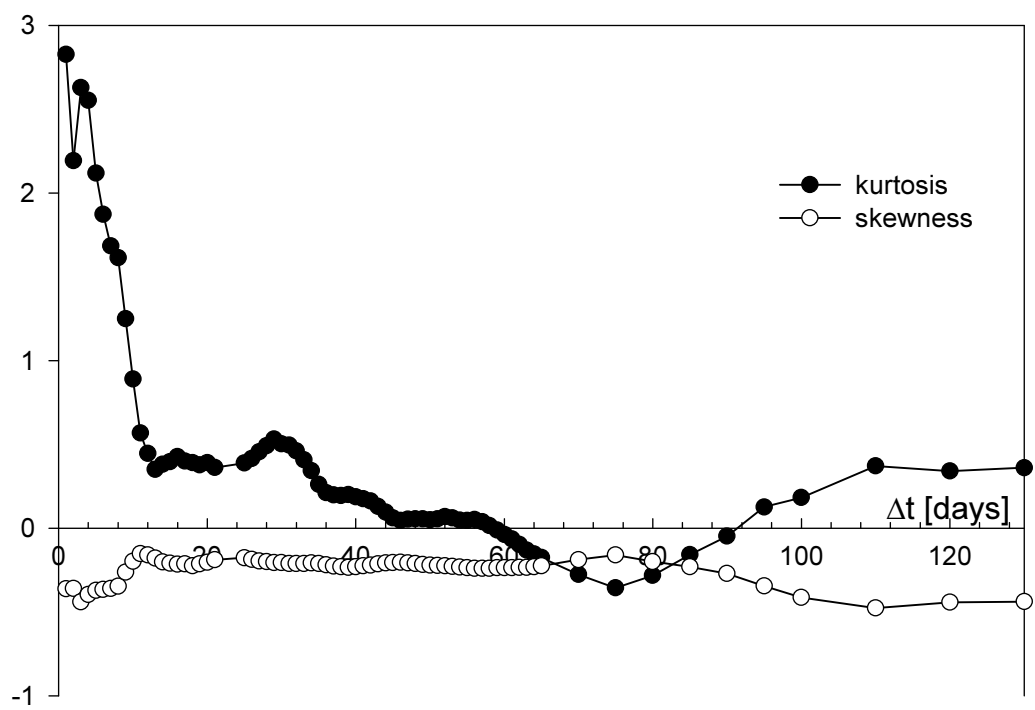
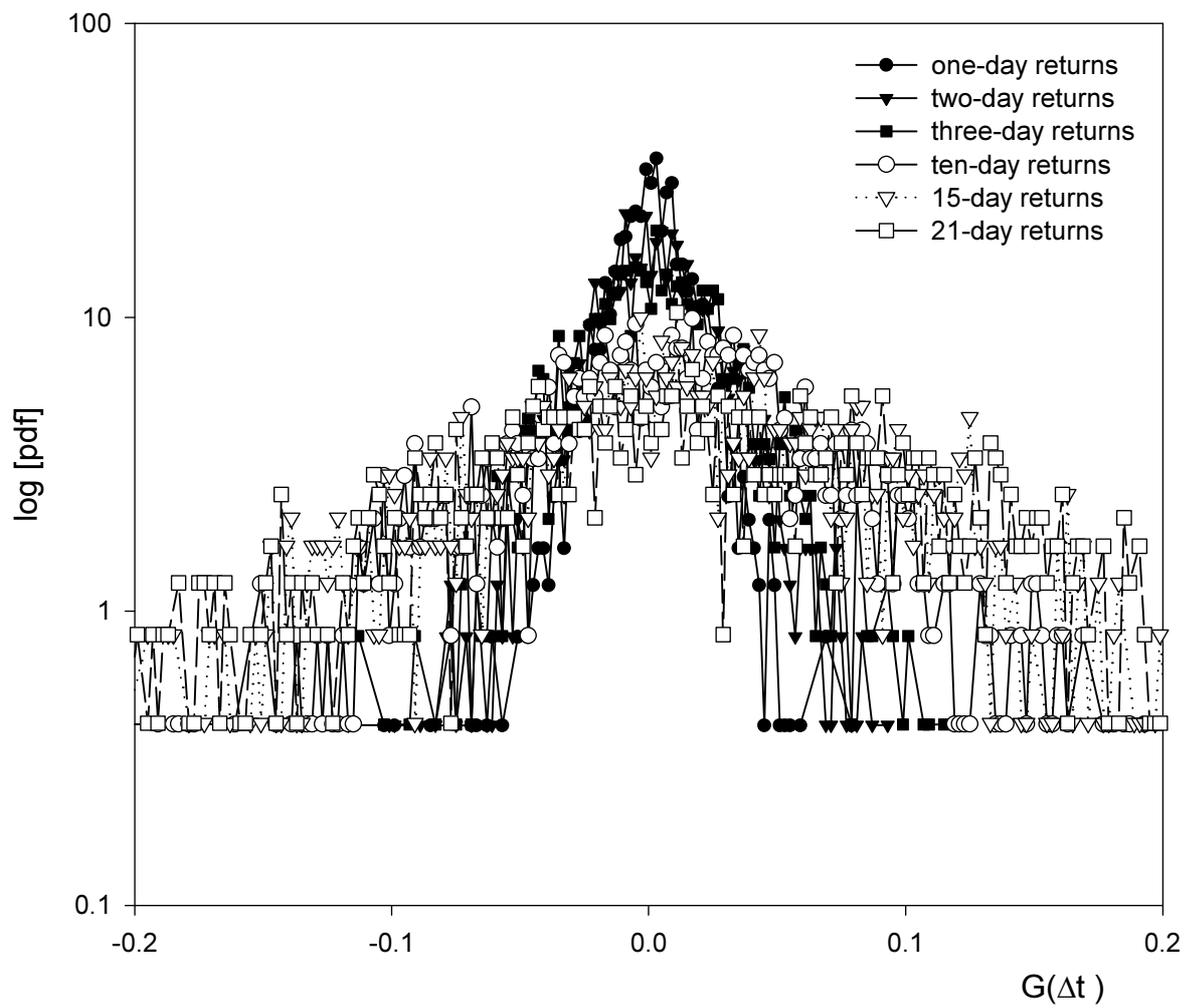
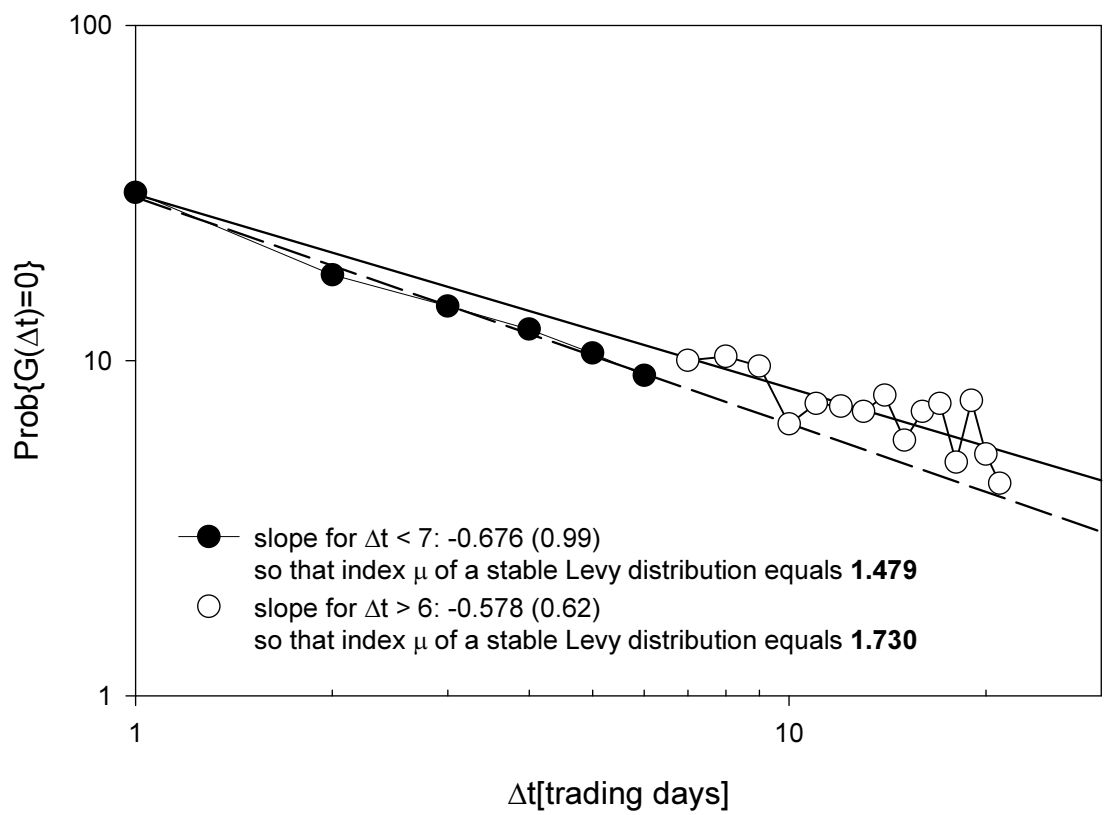
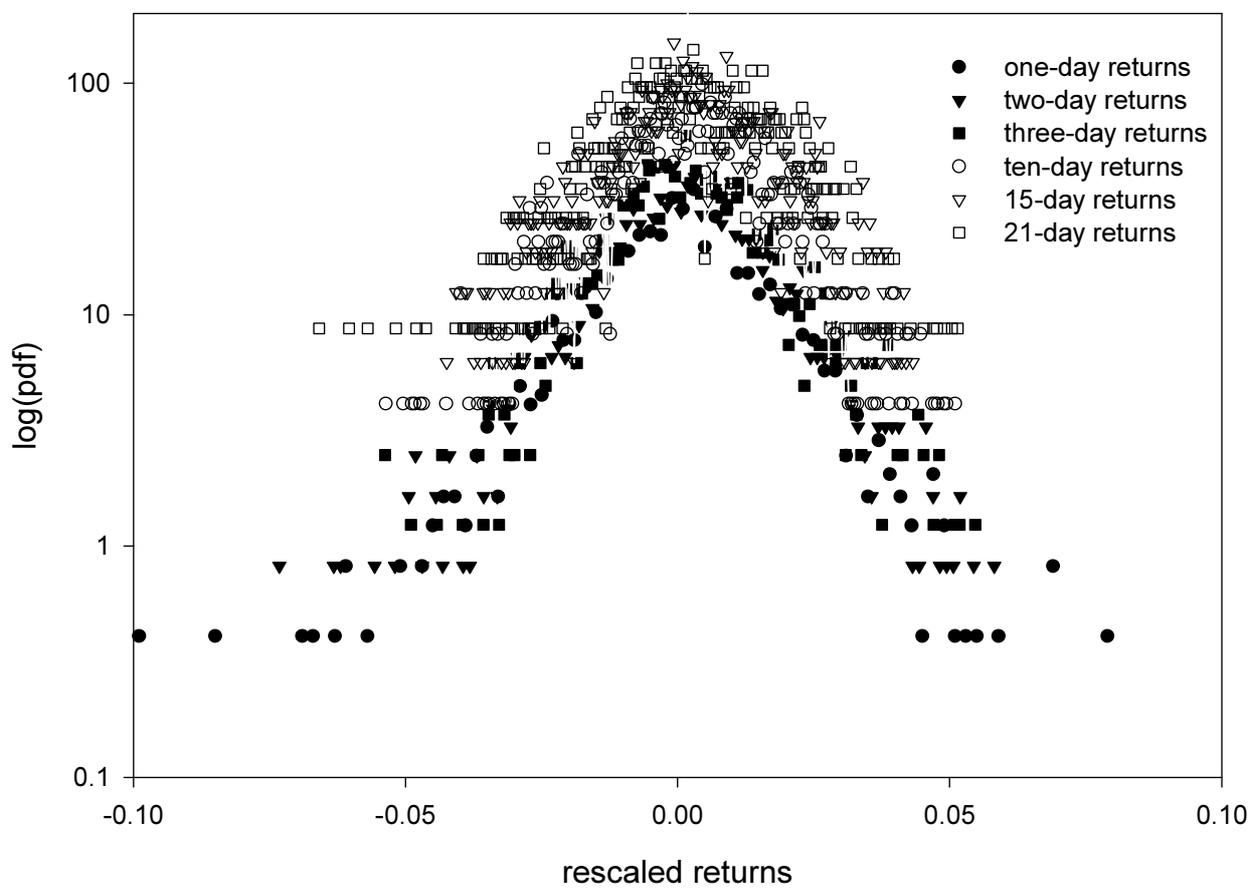
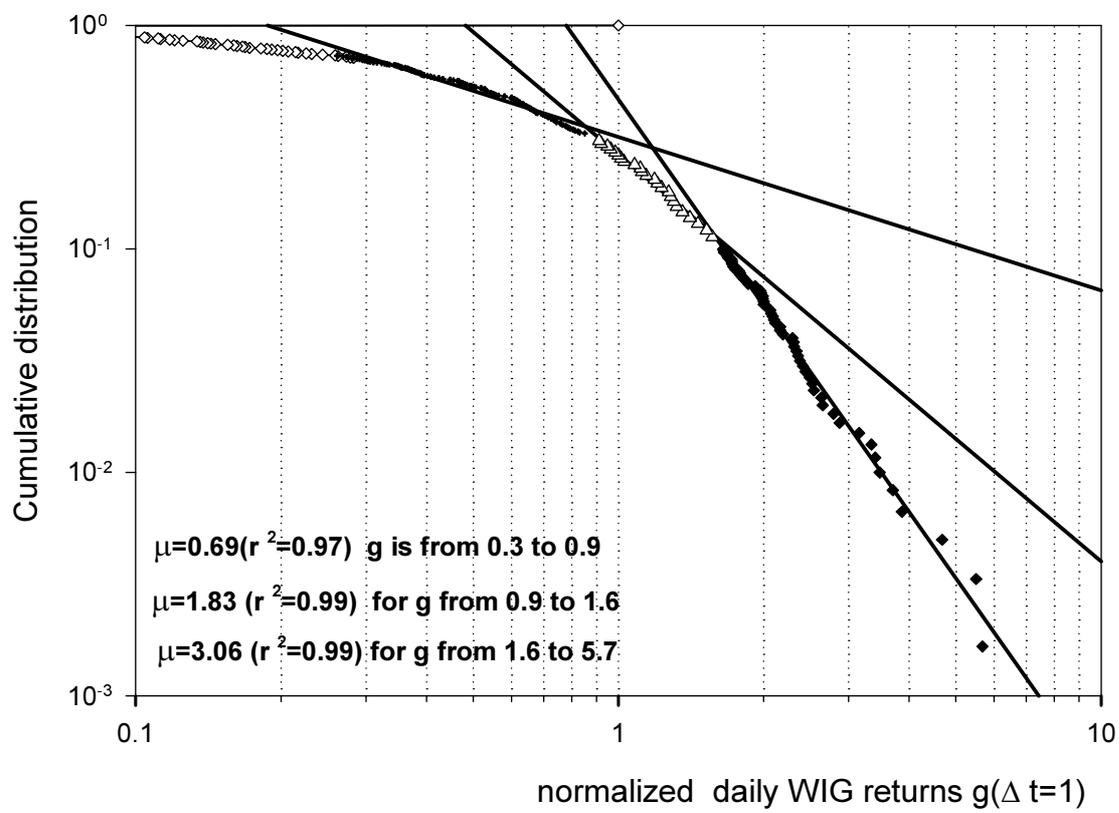
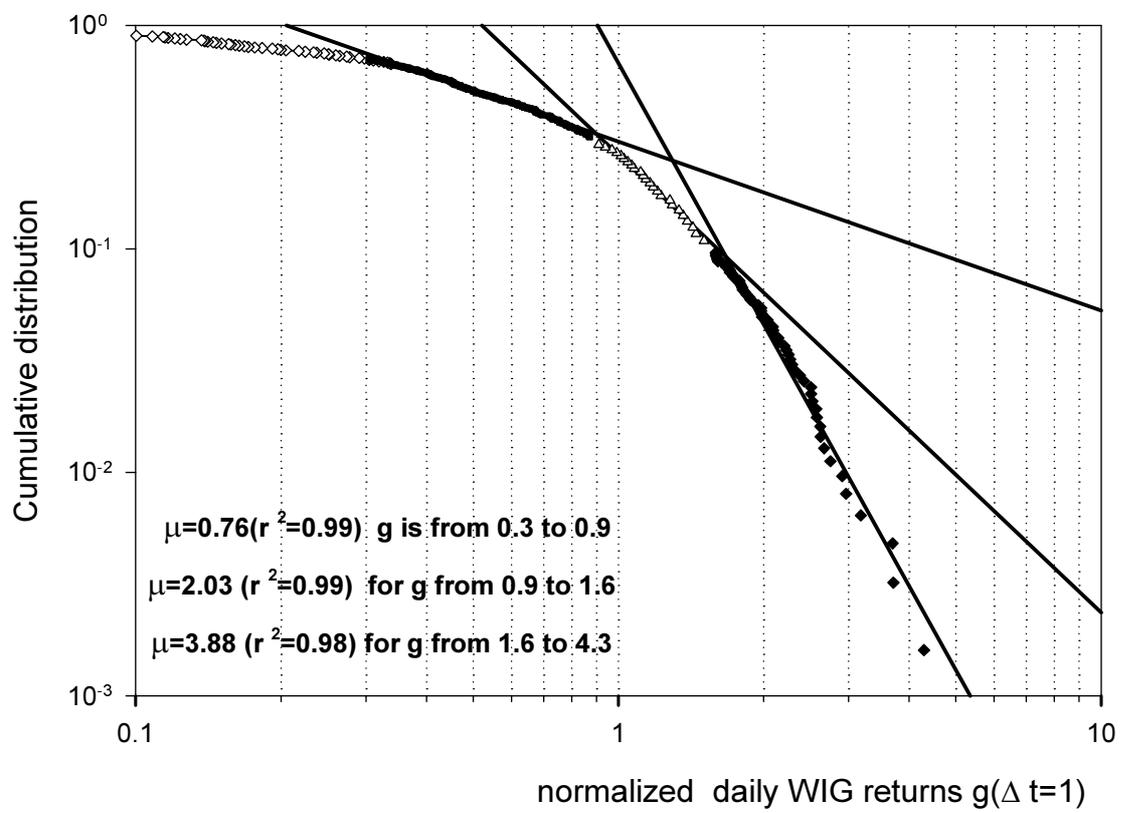
\end{document}